\documentclass[twoside]{article}
\usepackage{qic,epsfig}

\newcommand{\Tr}{\mathrm{Tr}}

\begin{document}
\setlength{\textheight}{8.0truein}    

\runninghead{An exact tensor network for the 3SAT problem}
            {Artur Garc\'ia-S\'aez and Jos\'e I. Latorre}
\normalsize\textlineskip

\thispagestyle{empty}
\setcounter{page}{1}


\vspace*{0.88truein}

\alphfootnote
\fpage{1}

\centerline{\bf
An exact tensor network for the 3SAT problem}
\vspace*{0.37truein}
\centerline{\footnotesize
Artur Garc\'ia-S\'aez and Jos\'e I. Latorre}
\vspace*{0.015truein}
\centerline{\footnotesize\it Dept. d'Estructura i Constituents de la Mat\`eria, Universitat de Barcelona }
\baselineskip=10pt
\centerline{\footnotesize\it 647 Diagonal, 08028 Barcelona, Spain}


\vspace*{0.21truein}

\abstracts{
We construct a tensor network that delivers an unnormalized 
quantum state whose coefficients are the solutions to a given
instance of 3SAT, an NP-complete problem. The tensor network contraction that
corresponds to the norm of the state counts the number of solutions
to the instance. It follows that exact contractions of this tensor network are
in the \#P-complete computational complexity class, thus believed to be a hard task. 
Furthermore, we show that for a 3SAT instance with $n$ bits, it is enough to
perform a polynomial number of contractions of the tensor network structure
associated to the computation of local observables to obtain one of the
explicit solutions to the problem, if any. 
Physical realization of a state
described by a generic tensor network is equivalent to finding
the satisfying assignment of a 3SAT instance and, consequently,
this experimental task is expected to be hard.
}{}{}

\vspace*{10pt}


\vspace*{1pt}\textlineskip    

\section{Introduction}
The rational to develop a quantum computer
follows from the fact that simulating quantum systems
in a classical computer is a hard task \cite{Feynman82}.
Nonetheless, an enormous progress has
been achieved to improve on the classical simulations
of relevant quantum problems using Monte Carlo
and Tensor Networks techniques. In particular,
networks of tensors can be used to represent the
coefficients of the wave function in a manner that
faithfully represents the entanglement of the quantum state.
The most relevant tensor networks geometries correspond
to one dimensional Matrix Product States (MPS \cite{Werner,Ostlund}), 
to higher dimensional Projected Entangled Pair States (PEPS \cite{VC04,PEPSrev}) 
and to Multiscale Renormalization Group Ansatz (MERA \cite{VidMERA}).
The limits for the efficiency of the tensor networks technology 
depends on the amount of entanglement pervading the system.

The main difficulty encountered in the application
of tensor networks to simulate quantum states corresponds
to computing observables, that is, to perform the
contraction of all the indices in the tensor structure.
Let us illustrate this point using the computation of
the properties of the ground state of a given Hamiltonian.
The tensor network approach needs two steps. First,
it is necessary to find the individual tensors in the
tensor network that give the approximation to the ground
state. This can be done using different strategies. 
A possibility is to use local Euclidean evolution, 
based on the idea of using the Trotter expansion for the
total Hamiltonian divided in local terms (see for instance
ITEBD for infinite spin chains \cite{VidITEBD}). Second, once the 
tensor network has been constructed, a contraction of the tensor
network is needed to perform the
computation of an observable. This contraction can be shown to
be efficient in one dimension but becomes unpractical
in higher dimensions due to the typical area law scaling of the
entropy found in any quantum system governed by local
interactions. Approximate contractions of tensor
networks do produce acceptable results, though 
the problem is worse and worse as the connectivity of
the system increases.

In order to assess rigorously the complexity class associated to
the problem of contracting arbitrary tensor networks, we here
present an exact representation of the solutions of a 3SAT problem
as an explicit tensor network. As we shall see, the tensor network
that delivers the solutions to a 3SAT instance can be constructed analytically
without the need of any minimization or training process. It is
then possible to relate the problem of contraction of tensor networks
to complexity classes associated to the 3SAT problem. As a matter
of fact, the 3SAT problem has been addressed previously using
quantum strategies \cite{Farhi}. In this context, the problem was analyzed using 
MPS in Ref. \cite{banuls,juanjo} as an approximation
technique to analyze the typical gap in an adiabatic
computational approach to NP-complete problems. Big instances of the problem can 
be solved, however, using classical statistichal methods \cite{parisi,survey}.

3SAT is an NP-complete problem \cite{cook,papad}, the complexity class 
of decision problems verifyable in polynomial time. Finding an explicit solution
to an NP problem defines the FNP complexity class, while counting the number of 
solutions to a polynomially computable function
defines the \#P complexity class. For instance, counting the solutions of an
instance of 3SAT is in \#P, thus believed to be a hard task. We show that the contraction 
of tensor networks for the 3SAT problem is a \#P problem, and
obtaining an explicit solution to the problem and reducing the bond dimension of 
the tensor network, are both in FNP.

This paper is organized as follows: in Section \ref{explicit} we present an explicit 
construction of the tensor network, used in Section \ref{complex} to analyze the complexity of 
the operations related to tensor network manipulation. In Section \ref{reduce} we show 
that reducing the virtual index of a tensor network is a hard problem. Finally,
we draw the conclusions in Section \ref{concl}.

\section{Explicit construction of a tensor network for 3SAT}\label{explicit}
Let us start by recalling the definition
of a 3SAT problem. An instance of 3SAT corresponds
to a decision problem, namely, determining whether there is
an assignment of a set of $n$ bits 
that can take values $x_i=0,1$, where
$i=1,\ldots,n$ such that a number of constraints are satisfied.
These constraints are defined by the set of 
$m$ clauses ($C_a, a=1,\ldots,m$), 
where each clause involves three bits and
rejects one of the eight possible assignments. To be more
explicit, let us consider a clause $C_a$ involving bits $i$, $j$ and $k$
and take the case where the assignment  
$(y_i,y_j,y_k)_a$ is excluded. This amounts to impose the boolean
restriction ${\overline {y_i\wedge
y_j\wedge y_k}}$. The 3SAT problem is defined as the problem
of deciding whether there exists a string of bits satisfying all the
clauses. As said, the 3SAT problem is NP-complete. 

\begin{figure} [htbp]
\centerline{\epsfig{file=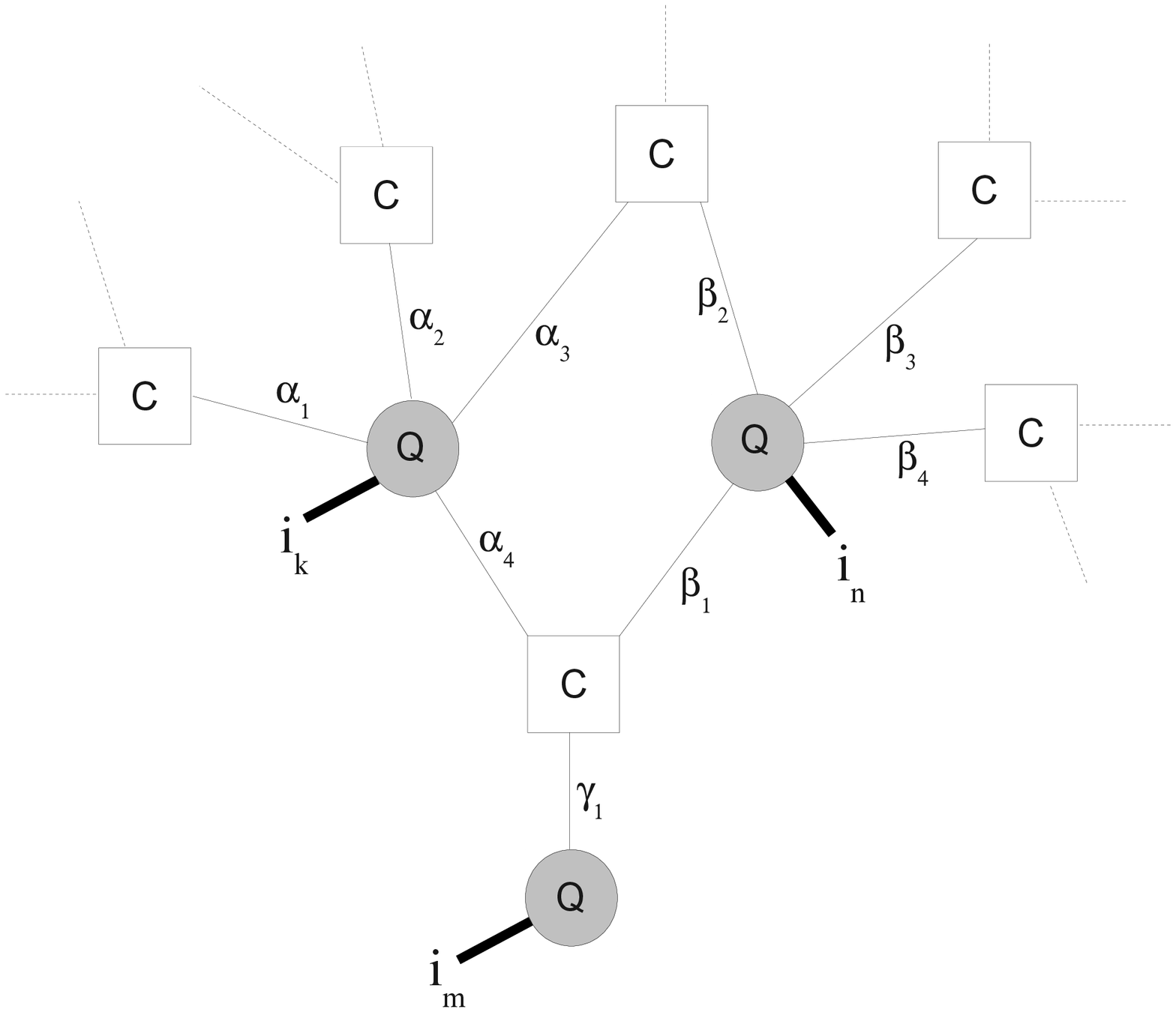, width=10cm}} 
\vspace*{13pt}
\fcaption{\label{fig1} We represent an instance of the 3SAT problem as a tensor network of bit tensors 
$Q^{[k],i_k}_{\alpha_1,\ldots,\alpha_r}$ (circles) and clause tensors $C^{[a]}_{\alpha\beta\gamma}$ (squares). 
Each clause tensor is connected to 3 bit tensors through the ancillary 
bonds $\alpha, \beta$ and $\gamma$. Bit tensors can be connected to an arbitrary number of clause tensors,
and possess a physical index $i_k$ related to the variable $x_k$. The contraction of this tensor network
provides a solution to the 3SAT instance.}
\end{figure}

Given a 3SAT instance, we construct a graph adapted to this 
specific instance under consideration. The graph contains two kind of
structures: one is associated to bits and another one to clauses. 
Each bit carries an open index related to $x_i$ and indices
linking the bit to the clauses related to it. Clauses have
no open index, they simply connect to the bits they relate.
An example of such a graph is presented in Fig.~\ref{fig1}. This kind 
of graph
naturally represents the connectivity of the 3SAT instance and
is common to many approaches to 3SAT as {\sl e.g.} the
belief propagation strategy \cite{parisi,survey}.
It is important to notice that not all instances of the 3SAT problem are hard. 
Considering a problem over $n$ variables and $m$ clauses, 
and defining $\alpha = m/n$, the problem is believed to be hard only in the critical range 
$3.42 < \alpha_c < 4.506$ \cite{parisi}. This constraint restricts the connectivity of tensor networks
descriving hard instances of 3SAT, which are represented by tensor networks only slightly connected. 
In our construction we only require $m+n$ tensors and $3m$ links. Notice that connectivity,
throught the planarity of the network, identifies as well hard instances of the Ising model
in higher dimensions \cite{istrail}.

Our next step is to construct an unnormalized quantum state
based on this structure. The idea is to associate a tensor 
to each element of the connectivity graph presented
above. Each bit is now associated to a tensor
$Q^{[k]i_k}_{\alpha_1,\ldots,\alpha_{r_k}}$ that describes qubit $k$.
The index $i_k$ can take values $0$ or $1$, that is the two possible states
of the qubit. The tensor for this qubit also carries ancillary
indices that we have represented with Greek letter $\alpha$'s
connecting to $r_k$ different clauses. Each qubit will connect
to a different number of clauses, so the total number
of links $r_k$ will depend on the qubit $k$ we are considering.
Every ancillary
index runs up to a maximum value $\chi$, $\alpha_i=1,\ldots,\chi$. 
Then, we further introduce a tensor for
each clause $C_a$, that we represent as $C^{[a]}_{\alpha\beta\gamma}$, where $a=1,\ldots,m$. 
Note that $\alpha$, $\beta$ and $\gamma$ are ancillary indices
that connect to three different qubits. The contraction of
all ancillary indices define our global unnormalized state 
\begin{equation}
|\psi(Q,C)\rangle=\sum_{i_1,\ldots,i_n}t^{i_1,\ldots,i_n}|i_1,\ldots,i_n\rangle
\end{equation}
where the coefficients $t^{i_1,\ldots,i_n}$
correspond to the contraction 
\begin{equation}
t^{i_1,\ldots,i_n}=\langle i_i,\ldots,i_n |\psi(Q,C)\rangle=\Tr\left[ {Q^{[1]i_1} \ldots Q^{[n]i_n}} {C^{[1]} \ldots C^{[m]}}\right],
\label{state}
\end{equation}
where the symbol $\Tr$ represents the full contraction of all the
ancillary indices that have been omitted in the above expression
for the sake of simplicity.

The problem we are addressing is how to assign the tensors $Q^{[k]i_k}$ and $C^{[a]}$ such that
the coefficients of the tensor $t$ are equal to $1$ for the solutions
to the 3SAT instance represented by the graph and $0$ otherwise. 
That is, 
\begin{equation}
t^{i_1,\ldots,i_n}=\left\{ \begin{array}{rl} 1,& (i_1,\ldots,i_n)=(y_1,\ldots,y_n)\\
0,& {\rm otherwise} \end{array} \right. .
\label{tensort}
\end{equation}
Then, it follows that the unnormalized state $|\psi\rangle$ corresponds to the equally 
weighted superposition of all the solutions to the 3SAT instance. 

It turns out that there is a very simple tensor network, with
only $\chi=2$, that carries all the solutions to a given 3SAT instance.
The basic idea comes from the fact that such a tensor can be thought
to be the ground state of a Hamiltonian made by the addition
of 3-body Hamiltonians, each one associated to one of the clauses
that define the instance,
\begin{equation}
H=\sum_{a=1}^m H_{a} .
\end{equation}
The 3-body Hamiltonian associated to the
clause $C_a$ involving qubits $i,j,k$ is constructed as
a projector on the wrong assignment to be rejected
$(y_i,y_j,y_k)_a$, 
\begin{equation}
H_a=|y^{[a]}_i,y^{[a]}_j,y^{[a]}_k\rangle \langle y^{[a]}_i,y^{[a]}_j,y^{[a]}_k| .
\end{equation}
That is, for a given clause $C_a$, its corresponding Hamiltonian
$H_a$ penalizes with one unit of energy the wrong
assignment to the clause and delivers a zero value
for the rest of the allowed assignments. 
The relevant point is that this Hamiltonian is diagonal in 
the computational basis, and all the pieces do commute
$\left[H_a,H_b\right]=0, \forall a,b=1,\ldots,m$, that is, the
Hamiltonian is frustration free, and local minimization provides a global minimum. 
Under some circumstances tensor networks {\it i.e.} PEPS have been identified 
as ground states of frustration-free 
Hamiltonians \cite{free}. Therefore, we can look for
a global tensor construction by just guaranteeing the correct 
local tensors clause by clause. Hence, we obtain our first and central
result:

\vspace*{12pt}
\noindent {\bf Result 1:} \emph{The solutions to an instance of the 3SAT problem can be 
represented explicitly as an unnormalized quantum state which is encoded in a tensor network with $\chi=2$.}
\vspace*{12pt}

The precise construction of the tensors that encode the solution
of the 3SAT instance corresponds to
\begin{equation}
Q^{[k]i_k}_{\alpha_1,\ldots,\alpha_r} = \left\{ \begin{array}{rl} 1,& \forall i_k: i_k=\alpha_1=\ldots=\alpha_r\\
0,& {\rm otherwise} \end{array} \right.
\label{tensorq}
\end{equation}
and
\begin{equation}
C^{[a]}_{\alpha\beta\gamma} = \left\{ \begin{array}{rl} 1, & (\alpha,\beta,\gamma)=(y_i,y_j,y_k)_a\\
0, & {\rm otherwise}  \end{array} \right.
\label{tensorc}
\end{equation}
All qubit physical indices $i_k$ are locked to the ancillary ones $\alpha_i$, and the values
of the qubit are neutral, taking the same value $1$ whether $i=1$ or $0$.
The tensor for the clauses reads the qubit value through the ancillary indices
and simply provide a $1$ for all the allowed assignments but a $0$ for
the rejected one. This structure is easily seen to minimize every $H_a$ and,
thus, provides the global ground state which is the even superposition of
all the global valid solutions to the set of clauses. That is, the contraction of
tensor network we have constructed does fulfill
Eq. \ref{tensort}.  

From our construction it is clear that there is a degeneracy 
of the tensor networks delivering the 
same state $|\psi\rangle$. It is possible to
multiply any coefficient of {\sl e.g.} a qubit tensor by
an arbitrary number, while dividing it in the associated clauses.
It is also possible to assign different lockings between the variable tensors
and the clause tensors so to produce the same solution.
This freedom is reminiscent of the so-called gauge symmetry
also present in other tensor network geometries \cite{PEPSrev}.
Local changes in the tensor values produce a continuum 
of tensor network constructions leading to the same solution.
Our choice in Eqs.~\ref{tensorq},~\ref{tensorc} can be
understood as a canonical form for this tensor structure.

The assignment in Eqs.~\ref{tensorq},~\ref{tensorc} is also found when a numerical 
Euclidean evolution is performed to the flat distribution $\left(|0\rangle+|1\rangle\right)^{\otimes n}$. 
The Euclidean minimization of the Hamiltonian $H$ produces the equally flat 
distribution of solutions. In some sense, this is surprising. 
It must be made clear that if a solution to the 3SAT 
instance exists, a simpler network with the qubit indices being locked to their
solution values will deliver the correct solution. 
From an initial random assignment to the qubits, we recover as well all the solutions with different weights.
In the case of a single satisfying assignment, a tensor of $\chi=1$ (a product state) also provides
the solution. Yet, the Euclidean evolution keeps restoring symmetry on
qubit assignments. As a consequence, the $\chi=2$ tensor network we have
presented is obtained as an infrared fixed point of an Euclidean evolution minimization
algorithm. 

\begin{figure} [htbp]
\centerline{\epsfig{file=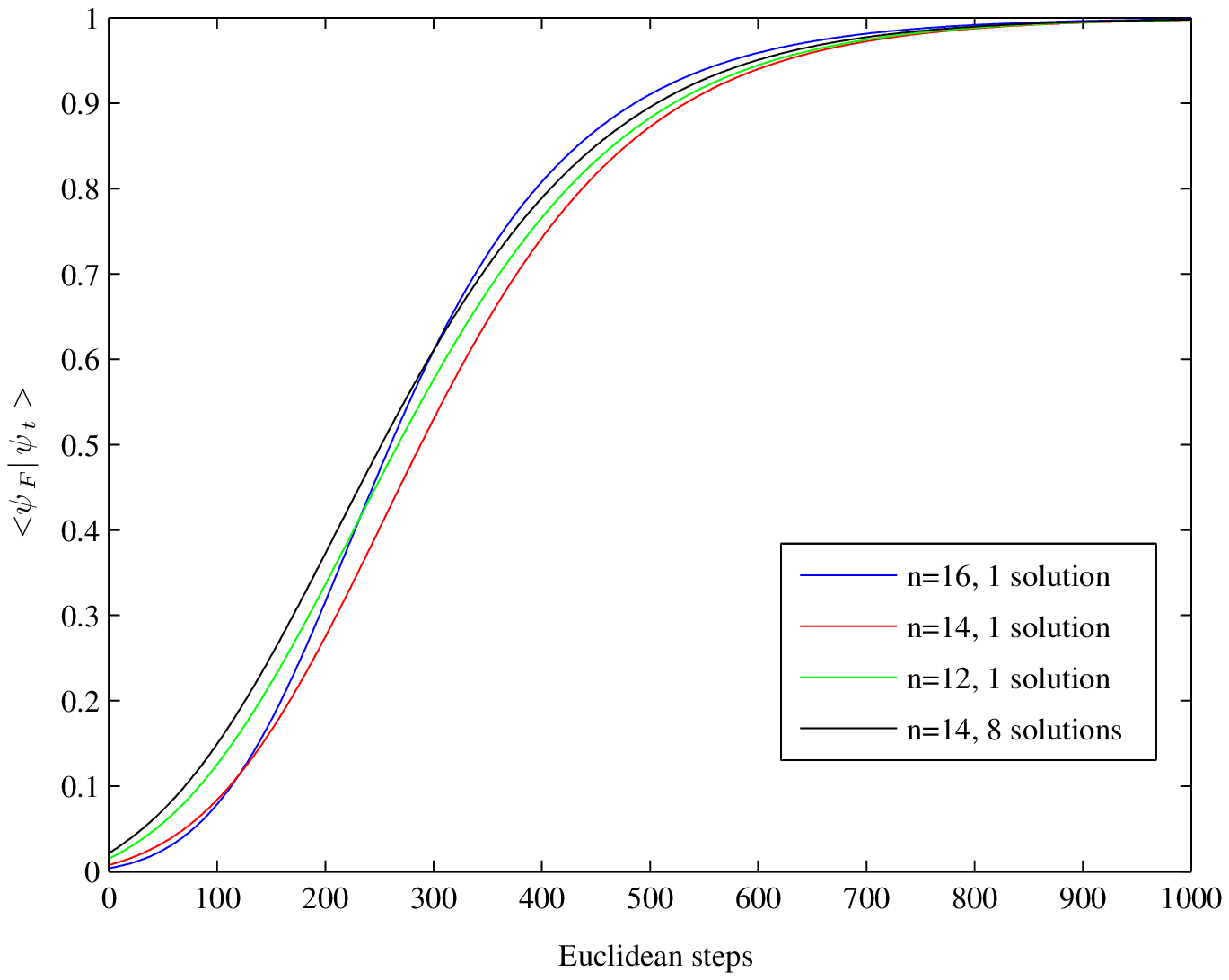, width=10cm}} 
\vspace*{13pt}
\fcaption{\label{fig2} For instances of 3SAT with $n=12,14,16$ variables we plot
the overlap of the evolved state $|\psi_t\rangle$ at time $t$ with the final 
state $|\psi_F\rangle$ that encodes the correct solution as described by 
Eqs.~\ref{tensorq},~\ref{tensorc}, along the Euclidean evolution.
We set the imaginary time step $\delta t = 0.001$. The initial overlap and the evolution
depends only on the number of variables and the number of solutions to the instance.}
\end{figure}

To further analyze the performance of the Euclidean evolution, we perform a numerical simulation
of the tensor network subject to the evolution operator. In Fig.\ref{fig2} we show the overlap between 
the final state encoding the solution and the evolved state along the imaginary time evolution. 
We fix $\delta t = 0.001$, resulting in $n=1000$ steps to provide the exact answer. 
The commutation between different $H_a$ ensures that every step of the computation
actually projects into the correct answer, identifying monotonically the right assignments already
in the first steps of the evolution. This is reflected on an identical evolution for instances 
with the same number of variables and solutions.

\section{Complexity class for the contraction of tensor networks}\label{complex}
\noindent
Note that the tensor network we have constructed is not normalized.
Indeed, if the 3SAT instance associated to this tensor has no solutions,
all the contractions leading to the computation of each $t^{i_1,\ldots,i_n}$
will produce a zero value. Such a state would have zero norm. In the case that
the 3-SAT instance has $p$ satisfying assignments we state the following:

\vspace*{12pt}
\noindent {\bf Result 2:} \emph{
Given the above tensor network that encodes the solutions to a 3SAT instance
in a state $|\psi(Q,C)\rangle$,
the norm $\langle\psi(Q,C)|\psi(Q,C)\rangle=p$
reads the number of solutions $p$ to the 3SAT instance. Hence, computing
the norm of the tensor network explicitly corresponds to a \#P-complete problem.}
\vspace*{12pt}

This result is derived from
the tensor network construction we have presented. 
It shows that a contraction of the tensor network counts solutions of
a 3SAT instance, which defines the \#3SAT 
problem. Consequently, the problem of contraction of arbitrary tensor
networks falls into the \#P-complete complexity class. This result
was derived in Ref. \cite{sharp} in the contex of contracting 
PEPS with bond dimension $\chi=2$.

Computing the number of solutions can be also reduced to the simpler case
of contracting $|\psi(Q,C)\rangle$ against a $\chi = 1$ tensor network representing the unnormalized state
$|\psi_{test}\rangle = (|0\rangle+|1\rangle)^{\otimes n}$. This contraction computes
$\langle\psi_{test}|\psi(Q,C)\rangle = \Tr\left[ {\tilde{Q}^{[1]} \ldots \tilde{Q}^{[n]}} C^{[1]} \ldots C^{[m]}\right] = p$,
with $\tilde{Q}^{[k]}= Q^{[k]0} + Q^{[k]1}$. That is, the simplest non-trivial contraction of a tensor network
with $\chi = 2$ with another with $\chi=1$ suffices to compute $p$. 
Note that this problem can also be seen as finding the overlap
of the 3SAT state $\vert \psi(Q,C)\rangle$ with the neutral product state
$| \psi_{test}\rangle=\sum_x \vert x\rangle$.

Let us emphasize further that the tensor network we have designed
only uses tensors with $\chi=2$ rank. This shows indirectly the enormous
representation power of tensor networks that use small rank individual
tensors. Important resources for quantum information processing can be 
constructed as well with tensor networks with $\chi=2$, such as the 
cluster state \cite{cluster} or the toric code state \cite{toric}. 
It was known that $\chi=2$ tensor network provide good approximations
to the ground states of relevant quantum interactions. PEPS with 
small rank for $\chi$ were 
shown to support area law scaling of entanglement and to
be able to describe a scale invariant theory where all
correlators decay polynomially \cite{toric}.

In our case, the situation is more radical due to the fact that
the connectivity is non-local. Then, the maximum entanglement support
offered by the tensor network can obey a volume law. 
More precisely, we can take a 3SAT instance and produce a bipartition
of qubits into two sets $A$ and $B$. Consider the case where
all qubits in $A$ are involved in a clause that further relates to
qubits in $B$. Our tensor network structure will then have 
at least $n/2$ ancillary indices connecting $A$ and $B$. 
Thus, the amount of entropy which can be supported by the
tensor network is $S_A\sim \log_2 \chi^\frac{n}{2}\sim n/2$, where $S_A$ is the von Neumann entropy of the
reduced density matrix for the subsystem $A$. This is the largest
possible entropy. Thus,  our tensor networks
can support  volume scaling for the entropy.
Numerical evidence support this necessity \cite{orus1,orus2}.

Let us now turn to the problem of finding an explicit solution to 
a 3SAT instance, which is a FNP problem. This can be achieved using contractions of
the tensor network we have constructed. From the fact that the 3SAT problem is self-reducible \cite{papad},
$n$ contractions of tensor networks deliver an actual solution, which leads to the following:

\vspace*{12pt}
\noindent {\bf Corollary:} \emph{An explicit solution to 
a 3SAT instance, encoded in a state $\vert \psi(Q,C)\rangle$,
can be obtained by computing a polynomial number of successive 
expectation values on that state.}
\vspace*{12pt}

We now turn to present a method to extract one of the
possible solutions, if any, of a 3SAT instance tensor network.
The basic idea is to perform a series of $n$ contractions as follows.
We first take any qubit $1$ and compute
\begin{equation}
\langle z_1\rangle=\frac{\langle \psi(Q,C)|z_1|\psi(Q,C)\rangle}
{\langle\psi(Q,C)|\psi(Q,C)\rangle} ,
\end{equation}
where $z_1=\frac{1}{2}(\sigma^z_1+I)$, an operator that reads $1$ or $0$ for
the elements of the basis of a qubit Hilbert space.
This operation is only defined if the norm of the state is different
from zero, that is, if there is at least one satisfying assignment
solving the 3SAT instance. The computation of this expectation value
amount to a new contraction of the tensor network that needs
exactly the same amount of computational cost as finding the norm
of the state. The result of the contraction can be any number between
$0$ or $1$. If the result is either one of the extremes, namely $0$ or $1$,
that means that the qubit must take such a value. The classical
bit associated to this qubit is completely fixed. 
In contrast, the case where the solution is not one of the two extreme values corresponds
to having more than one solution to the instance. 
In such a case, the state $|\psi(Q,C)\rangle$ is a superposition
of various solutions. The qubit which is
being analyzed cannot commit itself since it takes different values 
for those different solutions. Consequently, we can take any of the 
two possibilities and construct a new tensor network where 
$Q^{[k],i_k}_{\alpha_1,\ldots,\alpha_r}$ is fixed either to $0$ or $1$, as we prefer.
Both values will lead to a valid solution.
We can now proceed to a second qubit and repeat the operation. In this way,
$n$ contractions of tensor networks will suffice to deliver one of
the possible solutions to the 3SAT instance.

It should be noted that our approach requires a polynomial 
number of calls to \#P contractions. As a matter of fact, however, each call should 
only need to decide whether a solution exists, {\it i.e.} it would suffice to know if a certain assignment can be
satisfied at each step. Thus, each call would only amount to solving a NP problem.

\section{Reducing the rank of tensor networks}\label{reduce}
\noindent

A problem common to all the available techniques to make a contraction
of a tensor network corresponds to finding the best approximation
of a tensor network with large $\chi$ using another one with
a small $\chi$. For instance, in two dimensions, 
the contraction of a tensor network accumulates open ancillary
indices as dictated by the area law of entanglement. The contraction
of a tensor network needs a larger and larger effective rank $\chi$.
To solve this problem, different reductions can be tried. It is,
then, necessary to approximate large rank tensor networks by
small rank ones. 

The problem of reducing tensor networks is simpler in one dimension.
For MPS, the optimization problem of finding an appropriate reduction
of the ancillary indices requires polynomial resources \cite{MPDO}, though the global 
optimization problem can be extremely hard even in 1D \cite{eisert}. 
For the general of non-local tensor networks this task is out of reach.
This can be expressed as an evident statement.

\vspace*{12pt}
\noindent {\bf Result 3:} \emph{The reduction of a tensor network $\vert \psi(Q,C)\rangle$
that encodes a 3SAT instance
to a $\chi=1$ tensor network is equivalent to finding the solution to that instance.}
\vspace*{12pt}

This limitation can be deduced from our above results. Let us illustrate this 
point in the following way. As mentioned above, given an
instance of 3SAT with only one satisfying assignment,
we know of at least two tensor networks
that produce the correct identical state. One of them
is the tensor construction $| \psi(Q,C)\rangle$ with $\chi=2$ we have presented and a second
one is a tensor network with $\chi=1$ corresponding
to the product state. Both states are identical. Furthermore,
we know a priori the first construction since it is neutral
in all the tensor assignments. Then, the problem of finding
the best product state approximation to the $\chi=2$ tensor network,
that is finding 
\begin{equation}
\max_{|\psi_{\chi=1}\rangle}|\langle\psi_{\chi=1}|\psi_{\chi=2}\rangle|
\end{equation}
is equivalent to finding the solution to the instance. 
But, it was shown that finding the solution to the 3SAT instance is in FNP.
Thus, finding the optimal more economic tensor network construction
of a state given a more costly one is also in FNP.

Let us briefly discuss the relation of this result with practical applications
of tensor networks for condensed matter problems. The relevant point to note
is the two different categories of exact contractions {\it vs.} approximate solutions.
The reduction we have just discussed is an exact one. That is, given a $\chi=2$ tensor
network discribing the unique solution to a 3SAT problem, there is a reduction
problem to find the product state that delivers the same state.
On the other hand, in condensed matter problems it is interesting to know
whether a tensor network made of tensors of rank $\chi$ can be faitfully
approximated by a similar tensor network of lower rank $\chi'$. 
This is not an exact contraction and depends on the targeted accuracy
of the tensor representation. In general the larger $\chi$ is, the better
the approximation becomes. A quantitative analysis and characterization
of scaling properties of a tensor network approximation as a funtion
of $\chi$ can be found in Ref. \cite {chiscaling,chiscalingmoore}.
Further research devoted to determining the complexity of reducing
the rank of a tensor network for special cases with a lattice geometry,
as the case of Ising of Potts models, are known to be NP-complete 
\cite{bara,istrail,nest}.

\section{Conclusions}\label{concl}

Tensor networks provide a set of techniques to study by classical means interesting 
quantum states. These include universal resources for quantum computation, which can 
be expressed as tensor networks with  $\chi$ rank.
Using a simple construction that encodes the solution of a 3SAT problem in a tensor network,
$\vert \psi(Q,C)\rangle$,
we have shown that contracting arbitrary tensor networks, however loosely connected, 
may require exponential resources. From our construction, the number of solutions to
an instance of the 3SAT problem can be directly read from the  normalization $\langle\psi(Q,C)|  \psi(Q,C)\rangle$,
hence the difficulty of this task is equivalent to a contraction of
the network. The tensor network we have constructed is the ground state 
of a 3-local Hamiltonian. The hardness of finding its ground states implies
the difficulty of cooling such a spin system. 

Let us emphasize that the difficulty of finding a solution to a 3SAT instance
is not related to the construction of a tensor network for the state that encodes it.
There is no need to resort to complicate minimization techniques. Rather, 
the difficulty is attached to the contraction of the tensor network. 
The problem of counting solutions and of finding explicitly one of them
are solved with a polynomial number of contractions of the tensor network. 
It tourns out that a single contraction of the network counts the number of solution to the problem,
thus solving a \#P problem. The FNP problem of providing a single solution can be
solved with a controlled sequence of contractions.
Moreover, since the exact solution can be encoded in a $\chi=1$ tensor network,  
finding a reduction in rank for the tensor network is also FNP.

Since other hard problems reduce to the 3SAT problem addressed here, this result can
be extended easily to an explicit construction of solutions to other hard problems like
\emph{e.g.} Exact Cover.  
Our tensor network encodes the relation between clauses in a set of tensors, resembling the 
message passing structure of other approaches used to solve huge instances of 
the 3SAT problem \cite{parisi}. This provides an important link between techniques of both 
disciplines, suggesting the use of  schemes developed in
the study of 3SAT to improve the performance of contracting arbitrary
tensor networks in Quantum Mechanics. 

Let us make a further link between the above result
on computational complexity classes for the contraction
of tensor networks and the possibility of
constructing such tensor networks in the lab. In Ref. \cite{CiracSolano}
a practical protocol was presented to create Matrix Product States in
the laboratory. This idea has been further pursued in Ref. \cite{expPEPS}
in the context of PEPS. Our results show that the solution to
3SAT can be reduced to create the tensor network $|\psi(Q,C)\rangle$ we have presented
plus a set of $n$ measurements. The solution to the problem would be read from
the collapse of the state to one of the solutions which are
superposed in $|\psi(Q,C)\rangle$. As a consequence, it is expected that
creating the quantum state representing the tensor network we have presented 
would require an exponentially long time. This shows that only tensor networks 
that capture some special symmetry in the geometrical connectivity 
of the problem might be amenable to experimental realization.

\nonumsection{Acknowledgements}
\noindent
We thank R. Or\'us for discussions.
Financial support from MICINN (Spain), 
Grup consolidat (Generalitat de Catalunya), 
and QOIT Consolider-Ingenio 2010 is acknowledged.

\nonumsection{References}

\end{document}